\input{aipcheck.tex}

\def\selectedoptions{}
\ifx\empty\selectedoptions
  \def\selectedoptions{final}
\fi

\documentclass[
   \selectedoptions
  ]
  {aipproc}

\def\selectedlayoutstyle{6x9} 
\layoutstyle\selectedlayoutstyle

\SetInternalRegister\hbadness{8000} 

\newcommand\doingARLO[2][]{%
  \ifx\mmref\undefined #1\else #2\fi
}

\begin{document}
\null
{\flushright{FERMILAB-Conf-01/351-E}}

\title 
      []
      {Charm Lifetimes and Mixing}


\author{Harry W. K. Cheung\thanks{Talk presented at the 9th International
Symposium on Heavy Flavors, 
California Institute of Technology, Pasadena, Sept. 10-13, 2001.}}{
  address={Fermilab, P.O. Box 500, Batavia, IL 60510-0500},
  email={cheung@fnal.gov},
  thanks={}
}

\copyrightyear  {2001}

\begin{abstract}
A review of the latest results on charm lifetimes and $D$-mixing is presented.
The $e^+e^-$ collider experiments are now able to measure charm lifetimes quite
precisely, however comparisons with the latest results from fixed-target
experiments show that possible systematic effects could be evident. The new
$D$-mixing results from the $B$-factories have changed the picture that
is emerging. Although the new world averaged value of $y_{CP}$ 
is now
consistent with zero, there is still a very interesting
and favoured scenario if the
strong phase difference between the Doubly-Cabibbo-suppressed and 
the Cabibbo-flavoured
$D^0\rightarrow K\pi$ decay is large.
\end{abstract}

\date{\today}

\maketitle

\section{Motivation for the Study of Charm Lifetimes}

The study of charm lifetimes is essentially a study of strong interactions
\cite{Reference:bellinibigi},
and in particular provide a test of the theoretically challenged part of the
Standard Model, namely non-perturbative QCD. It is hoped that experimental
results in charm lifetimes (possibly combined with other charm results),
can give some guidance as to what is needed to theoretically describe
strong interactions at all energy scales. This is important not only to
improve our theoretical understanding of strong interactions, but also
because the theoretical tools used to calculate lifetimes are the same or
similar to those used in other areas, for example to extract $V_{cs}$ and
$V_{cd}$ in charm decays; to calculate the $b$-particle lifetimes; and
to extract other Standard Model parameters or decay constants in heavy
flavor physics.

The other motivation is more mundane. Theoretical calculations are used
to calculate decay rates whereas experimentally one measures branching
fractions. One needs precise particle lifetimes to convert measured
branching ratios to decay rates so one can compare to theory and to extract
Standard Model parameters. Lifetimes are also important as an experimental
tool since the correctness of the measured lifetime will test techniques or
probe systematic effects in other areas where lifetimes and lifetime resolution
is important. For example in $D$ and $B_s$ $\Delta m$ mixing
measurements or in $\Delta\Gamma$ measurements for $D$ and $B$ mesons.

\section{Comparison of Experiments}

Table \ref{tb_experiments}\ shows a comparison of the experiments for which
new results in charm lifetimes and mixing have been presented recently.
These include both fixed-target experiments and experiments at
$e^+e^-$ colliders. This is significant as the two types of experiments
are quite different and will thus have different systematics. Therefore
a comparison of the results between the two types of experiments will be
an important check of any systematic effects.

\begin{table}
\begin{tabular}{l|c|c|c|c|c|c}
\hline
  & \tablehead{3}{c}{b}{Fixed Target} 
  & \tablehead{3}{c}{b}{$e^+e^-$ Collider} \\ \hline
Experiment & E791  & SELEX  & FOCUS & CLEO  & BaBar & BELLE \\
Beam       &\multicolumn{2}{c|}{hadronic} & photon &
\multicolumn{3}{c}{off-resonance $e^+e^-$} \\
Charm      & $\sim 10^5$ & $\sim 10^4$ & \multicolumn{2}{c|}{$\sim 10^6$} &
\multicolumn{2}{c}{$>10^6$}\\
$\sigma_t$ (fs) & $\sim 40$ & $\sim 20$ & $\sim 40$ & $\sim 140$ &
\multicolumn{2}{c}{$\sim 160$}\\
Method     & \multicolumn{3}{c|}{Uses vertex detachment cut}
           & \multicolumn{3}{c}{Needs average IP position} \\
           & \multicolumn{3}{c|}{Uses 3-D decay length}
           & \multicolumn{3}{c}{Uses $\sim$2-D decay length}\\
\hline
\end{tabular}
\caption{Comparison of experiments with recent results on charm lifetimes
and mixing}
\label{tb_experiments}
\end{table}

\begin{figure}
  \includegraphics[height=0.8in]{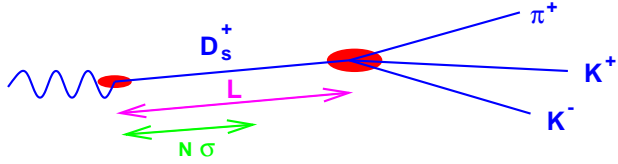}
  \caption{Illustration of decay length $L$ and vertex separation}
\label{fg_decay}
\end{figure}

\begin{figure}
  \includegraphics[height=1.3in,width=.24\textwidth]{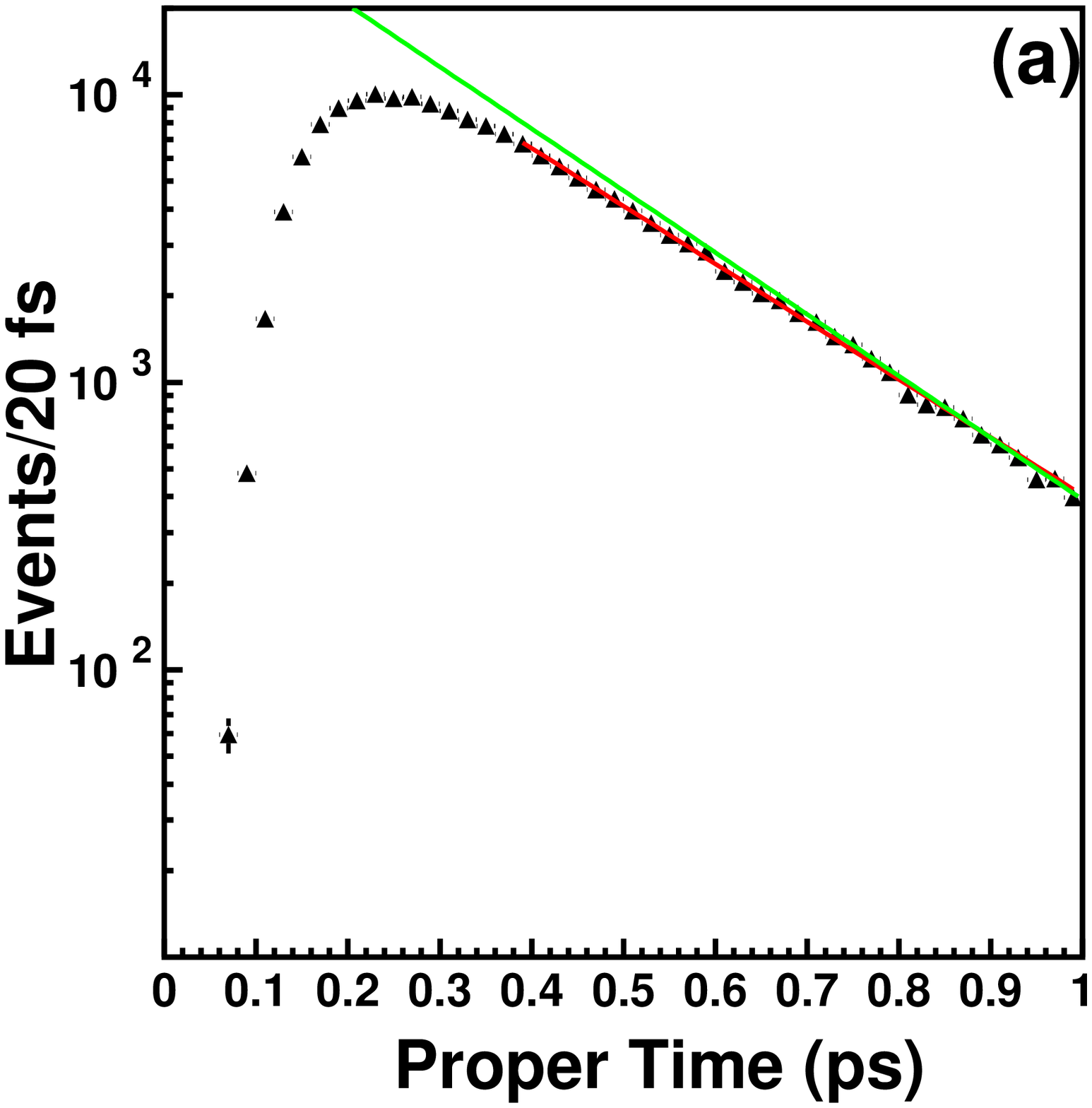}
  \includegraphics[height=1.3in,width=.24\textwidth]{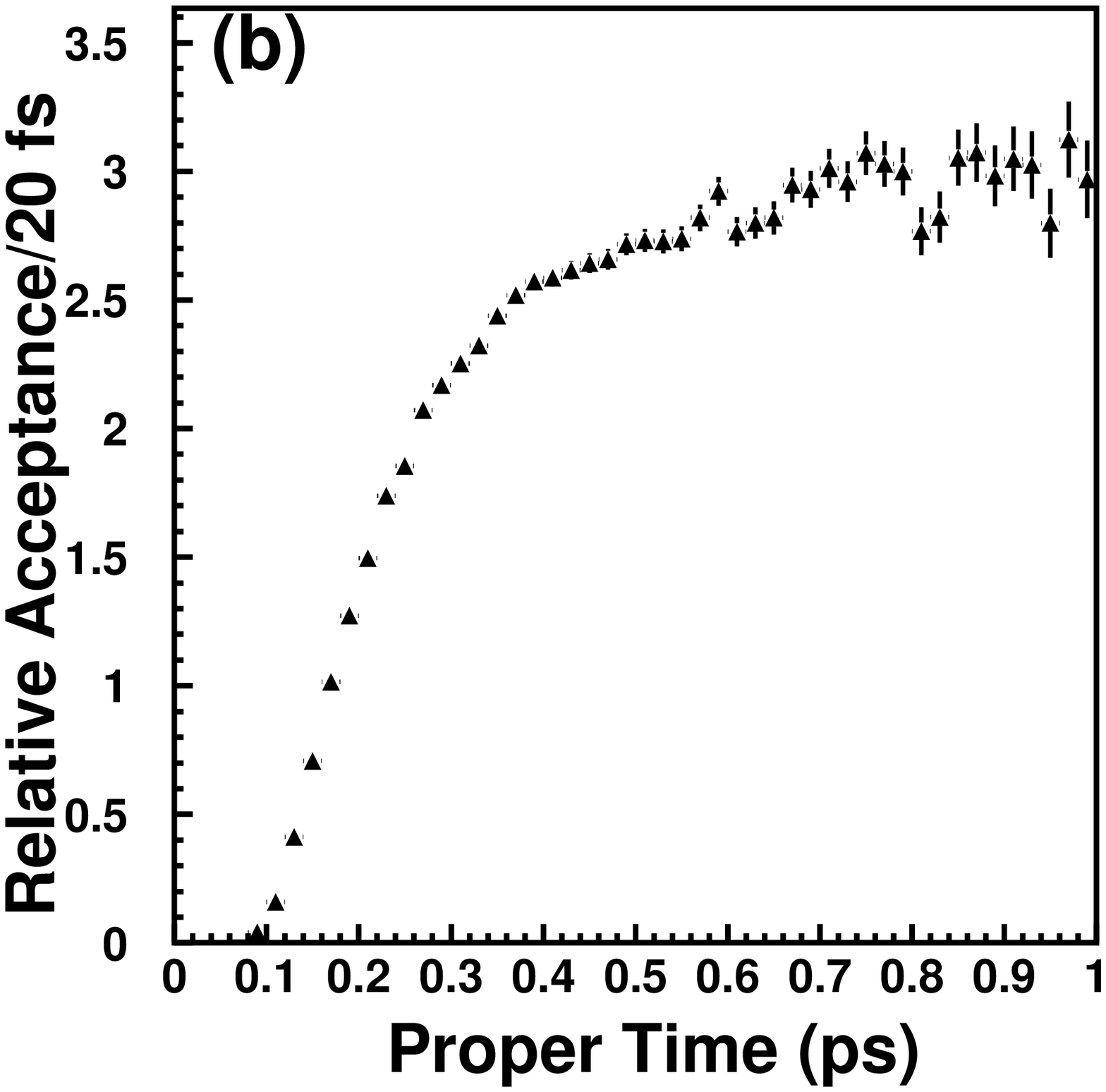}
  \includegraphics[height=1.3in,width=.24\textwidth]{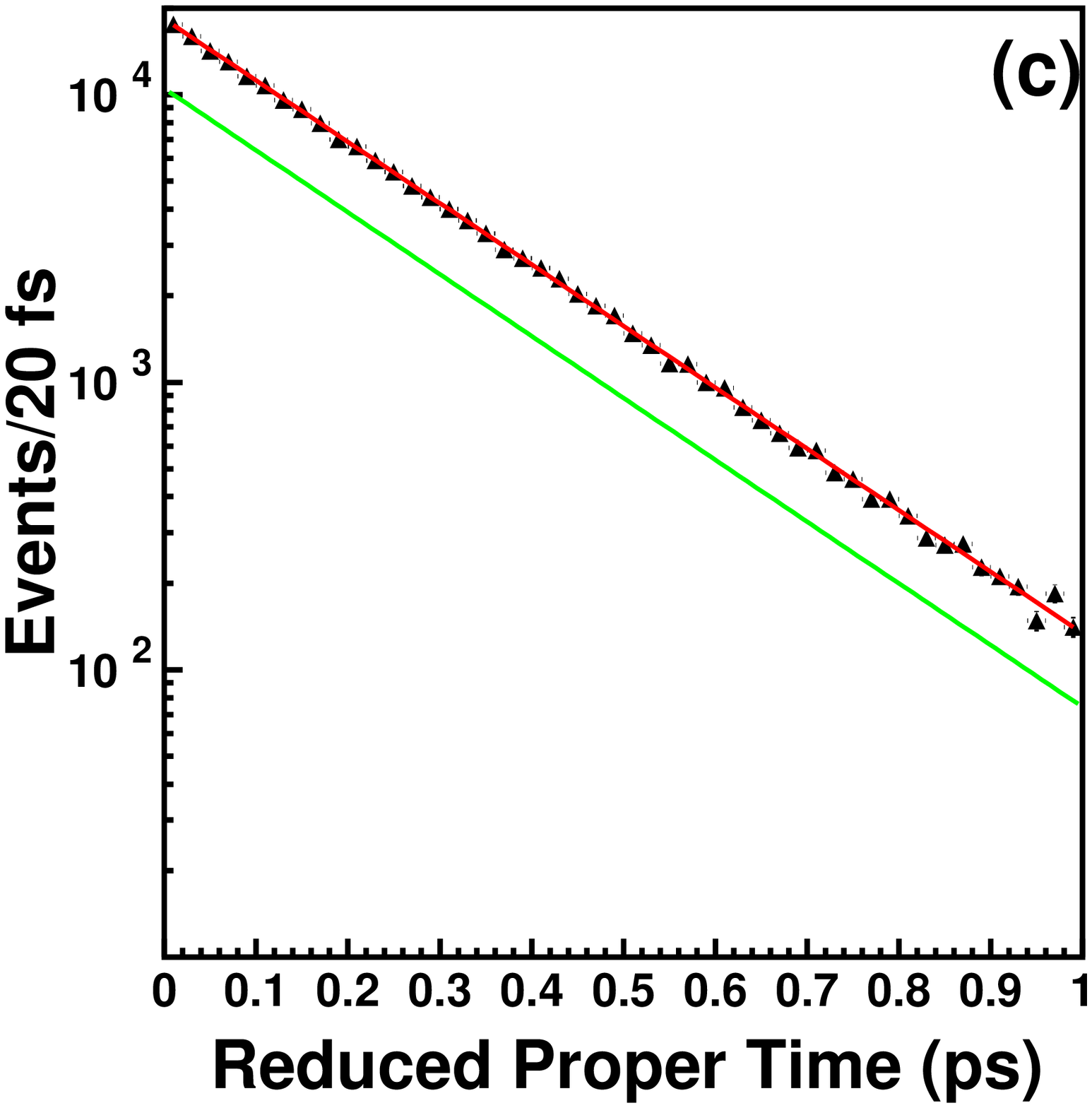}
  \includegraphics[height=1.3in,width=.24\textwidth]{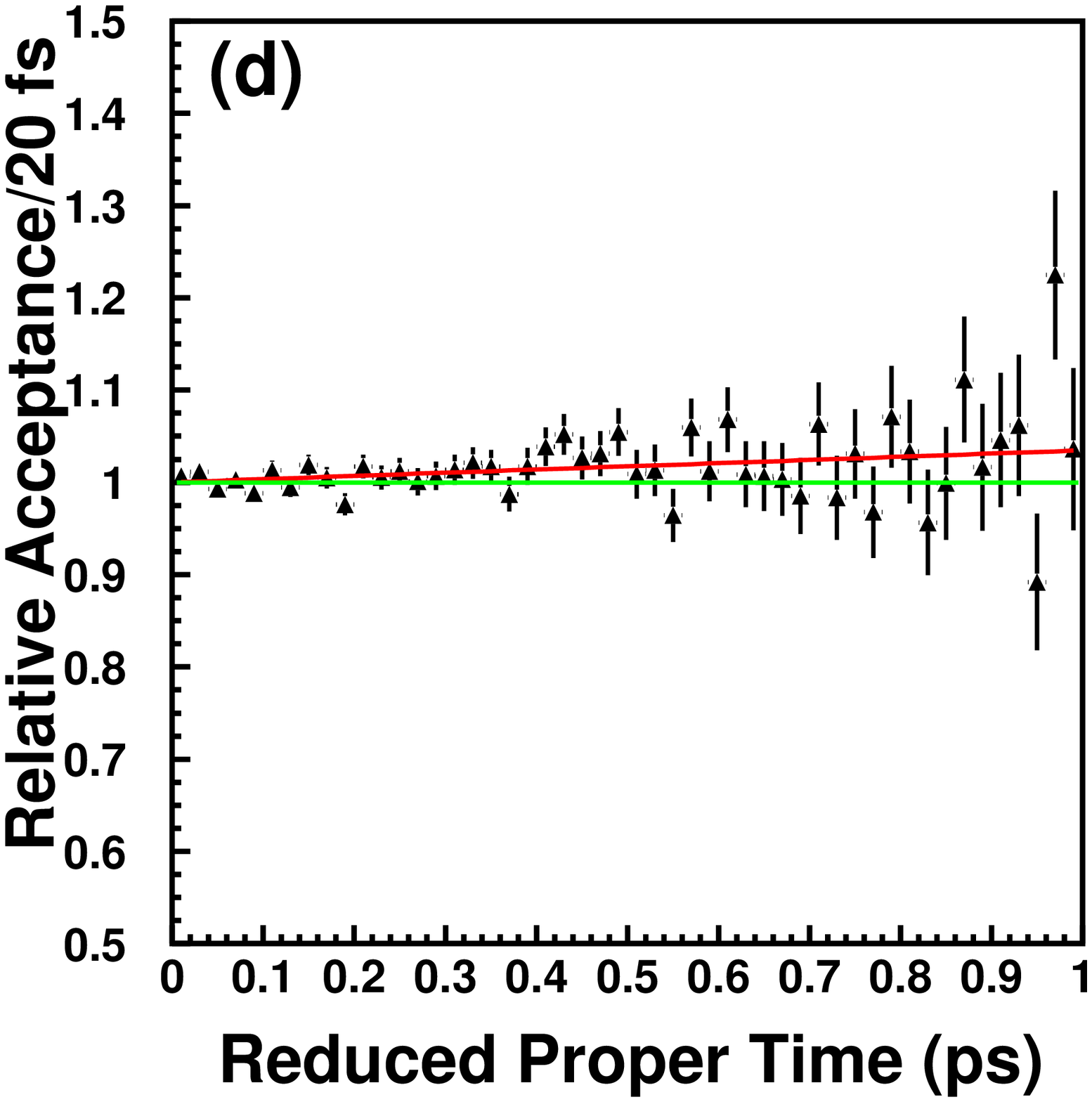}
  \caption{(a) Proper time distribution for a MC sample when a 
vertex separation requirement is made to reduce backgrounds; (b)
resulting acceptance correction function; (c) reduced proper time
distribution showing an exponential where the offset line shows the
slope of the true generated lifetime; (d) acceptance as a function
of reduced proper time.}
\label{fg_ft1}
\end{figure}

Typically fixed-target experiments have excellent vertex and proper time
resolutions in 3-dimensions but
large (non-charm) backgrounds. These backgrounds are
eliminated by selecting decay vertices that are well separated
from the production vertex, {\em e.g.} $L>N\sigma_L$ (see Fig.\ref{fg_decay}),
and optionally also outside of target material. This means that short-lived
decays are preferentially eliminated and the proper time distribution would
look like that given in Fig.\ref{fg_ft1}(a) requiring large non-uniform
acceptance corrections as illustrated in Fig.\ref{fg_ft1}(b).
This problem is avoided by using the reduced proper time,
$t^{\prime}=(L-N\sigma_L)/\beta\gamma c$, which starts the clock at the
minimum allowed proper time. The lifetime follows the same exponential
wherever one chooses to start, so the reduced proper time distribution
will follow an exponential with the true lifetime. This is illustrated in
Figures \ref{fg_decay}(c) and (d). The acceptance correction 
obtained using Monte Carlo simulations can be checked with data
by using $K_s^0$ decays reconstructed with the vertex detector as these
have a well measured lifetime. Absorption corrections are typically
small and can similarly be checked in data.

In fixed-target experiments there is usually a compromise between systematics 
due to backgrounds and systematics due to the acceptance correction.
The latter can become larger if one uses other types of vertexing
cuts to eliminate more background at the expense of introducing a (larger)
lifetime dependence in the acceptance. 

\begin{figure}
  \includegraphics[height=1.3in,width=.49\textwidth]{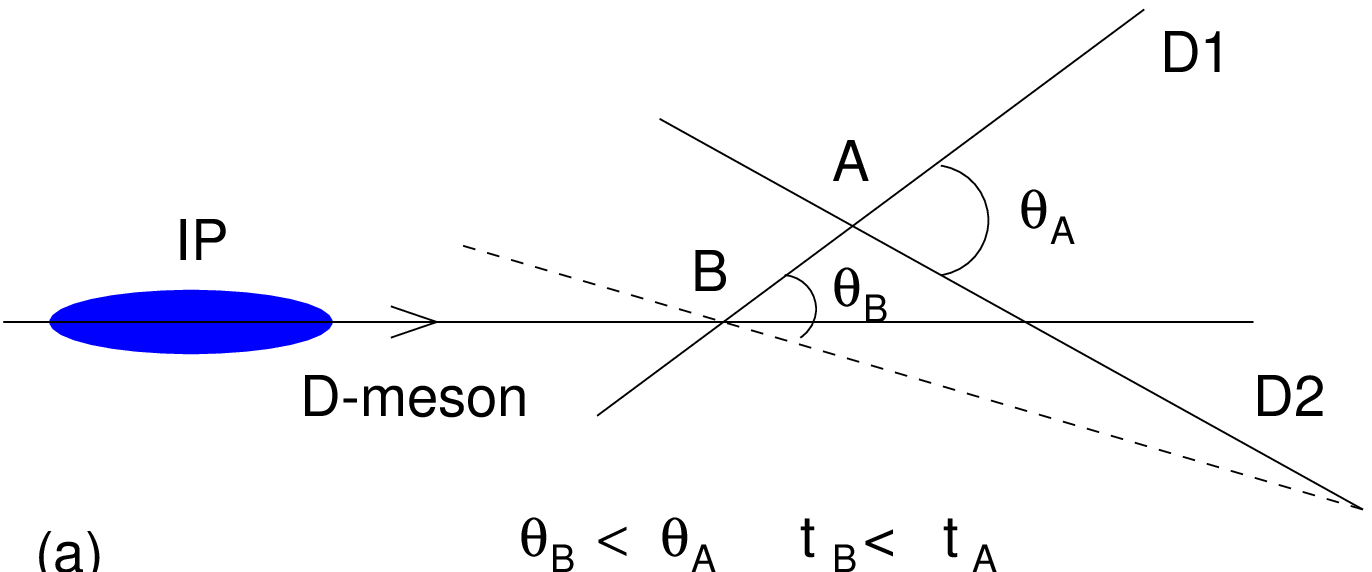}
  \includegraphics[height=1.3in,width=.24\textwidth]{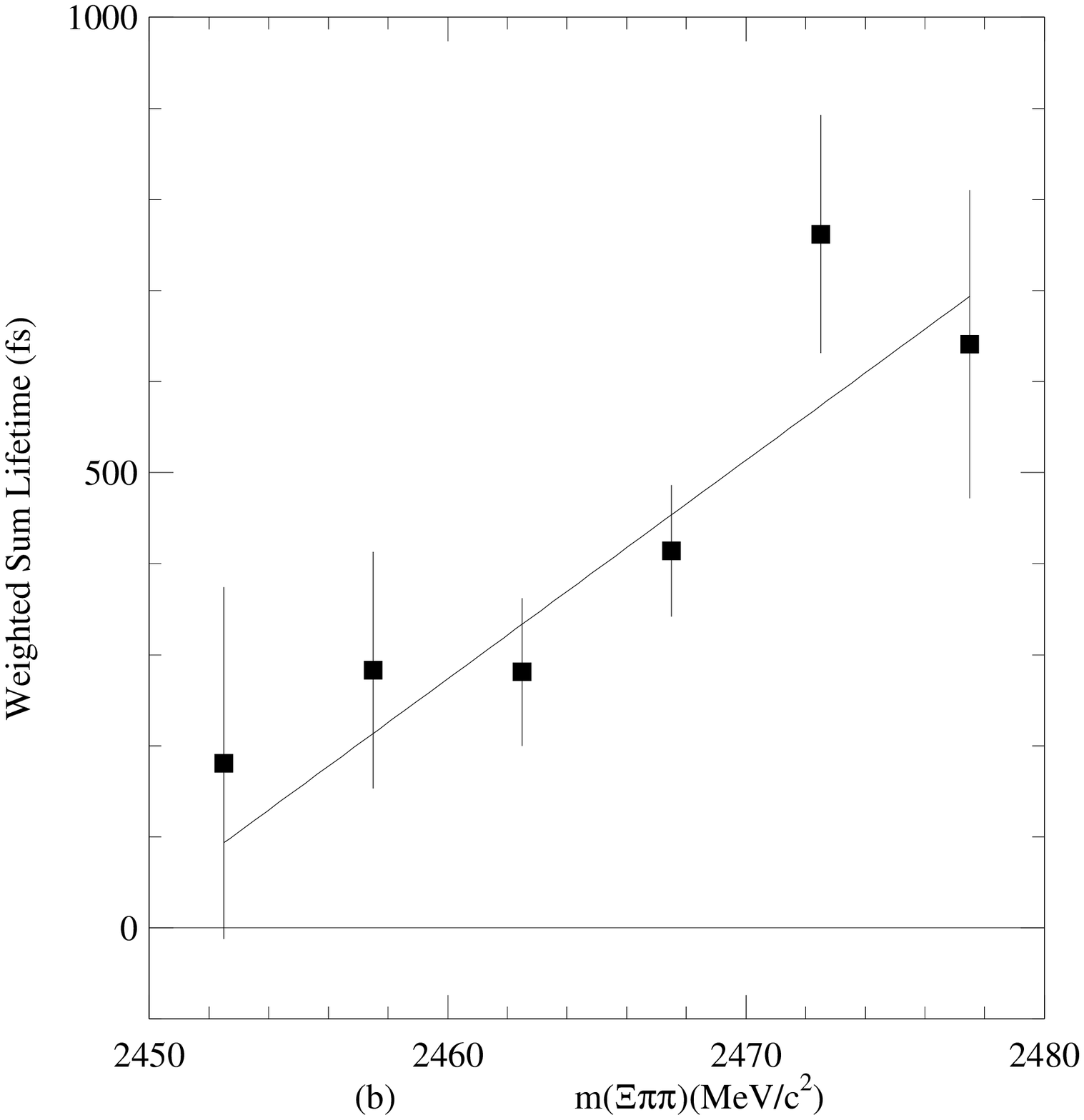}
  \includegraphics[height=1.3in,width=.24\textwidth]{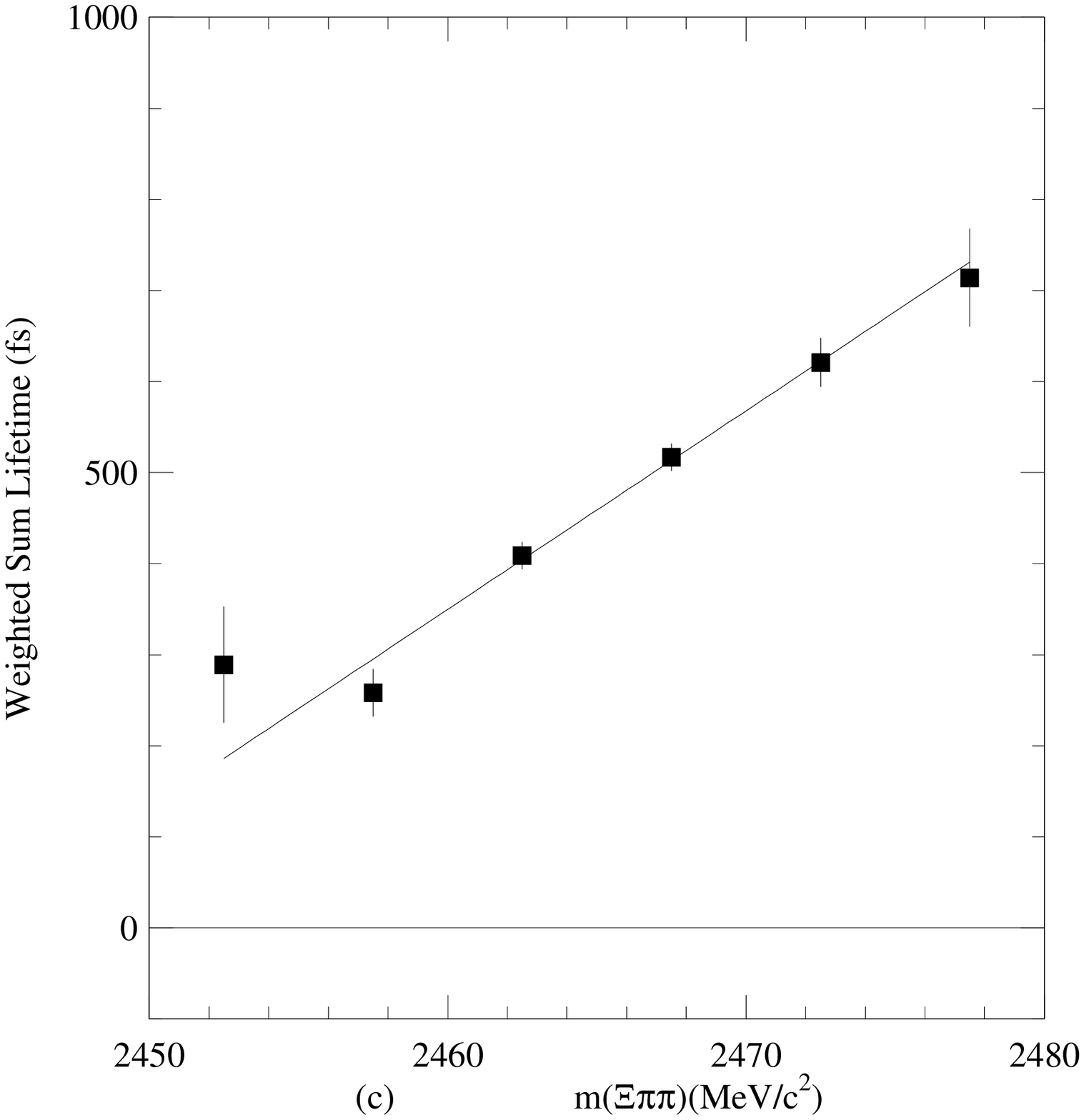}
  \caption{(a) Illustration of altering a track slope so that track $D2$
is at intersection point $B$ instead of the unbiased initial
reconstruction point $A$; (b) Correlation between mass and lifetime as
seen by CLEO in data; and (c) in MC.}
\label{fg_ft2}
\end{figure}

In contrast, the $e^+e^-$ experiments operating near $B$ threshold have
much less background but have poorer vertex resolutions and hence 
poorer proper time resolutions. The average interaction point is normally
used for constraining the location of the production point but its position is
usually only known well in 2-dimensions or even only 1-dimension.
The proper time resolution can be large compared to the charm particle
lifetime under study.
Due to proper time smearing from the poor resolution 
one has to take into account this smearing on an event-by-event basis.
This necessarily requires a more complicated event-by-event likelihood analysis
where one has to parameterize the time and mass resolutions as well as
background lifetime distributions. These resolutions can be known well
as they can be obtained from
data but they are not usually parameterizable by a simple function.
The resolution can sometimes be improved by using additional constraints
like forcing the reconstructed decay secondaries to come from a single
point or using the average IP position. However this can also lead to fit
biases and subsequent corrections. An example of this is illustrated in
Fig.\ref{fg_ft2}(a) where constraining a decay track to come from point
$B$ instead of $A$ will decrease the openning angle and hence also the
reconstructed mass, it would also decrease the lifetime. This produces
a correlation between the reconstructed mass and lifetime as seen in
CLEO for $\Xi_c^+\rightarrow \Xi^-\pi^+\pi^+$
decays in both data, Fig.\ref{fg_ft2}(b), and MC Fig.\ref{fg_ft2}(c)
\cite{Reference:cleoxicpplt}.
Systematic concerns are therefore usually 
related to the fit method, resolutions
and fit biases. 

\section{Results on Lifetimes}

\begin{figure}
  \includegraphics[height=1.5in,width=.33\textwidth]{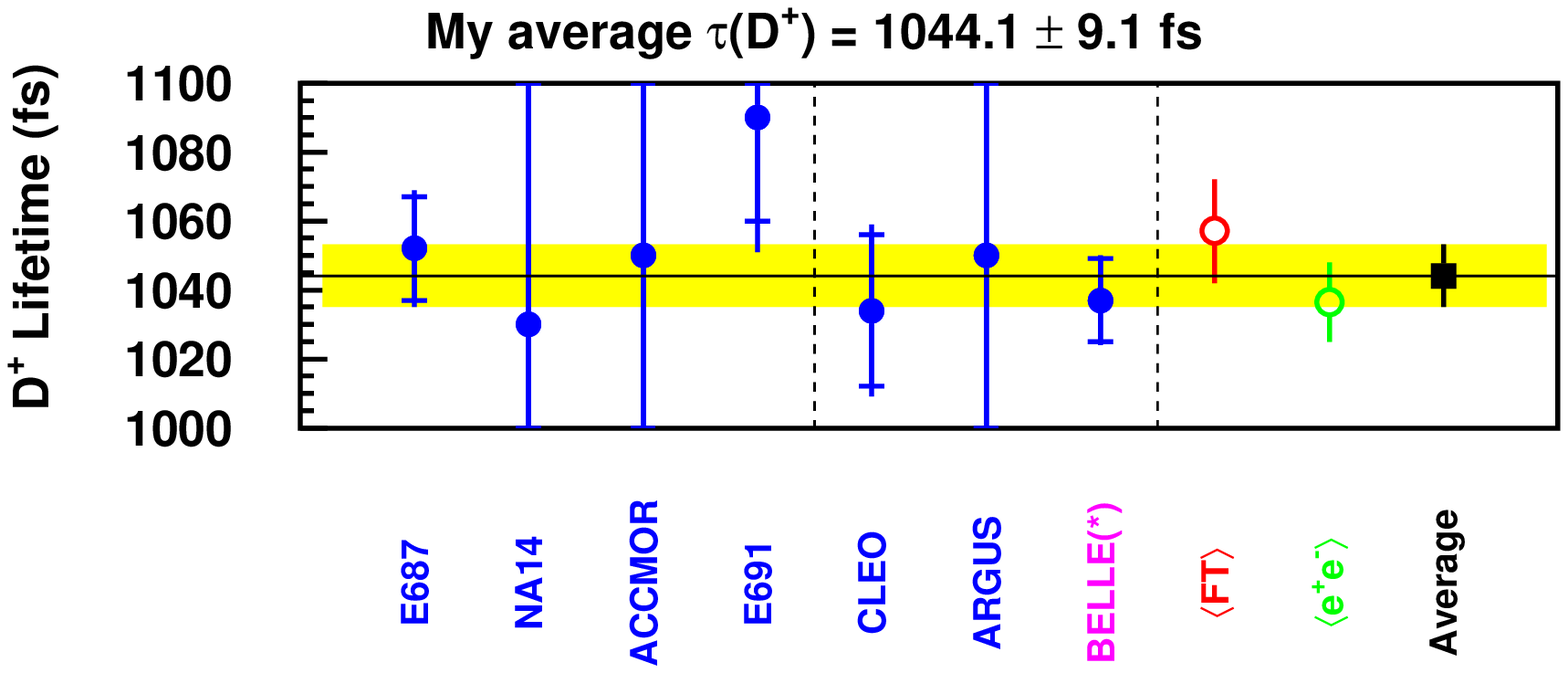}
  \includegraphics[height=1.5in,width=.33\textwidth]{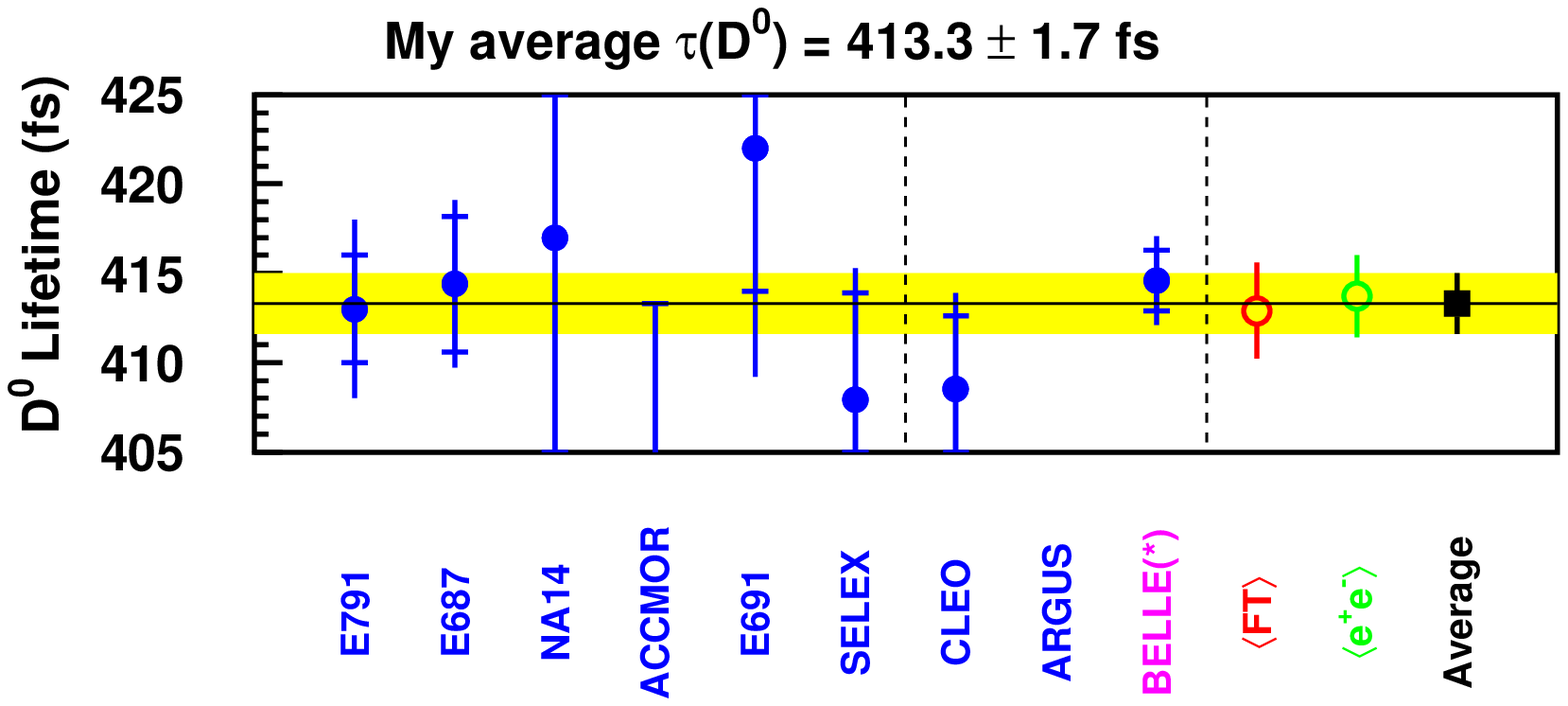}
  \includegraphics[height=1.5in,width=.33\textwidth]{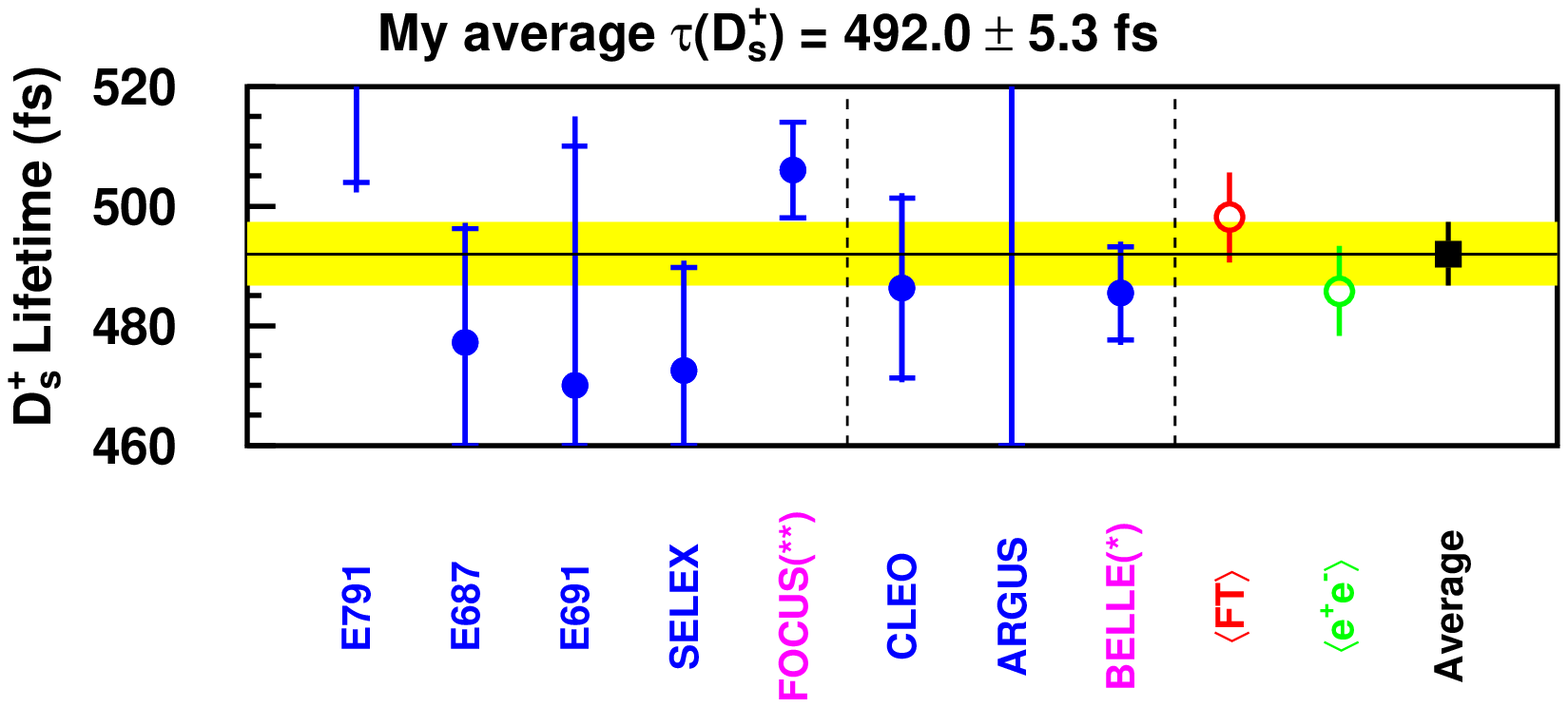}
  \caption{Summaries of recent $D$-meson lifetime measurements
and their averages.}
\label{fg_dltsummary}
\end{figure}

Both BaBar and BELLE would like to demonstrate that they have excellent
understanding of their vertexing and lifetime resolutions by measuring
the lifetimes of the charm particles. In fact they already have enough
data to get charm lifetimes with a precision comparable to the current
world averages. BELLE has led the way with preliminary
lifetime measurements for
the $D^0$, $D^+$ and $D_s^+$ mesons \cite{Reference:bellelt}.
A summary of the most recent charm lifetime results are given in
Fig.\ref{fg_dltsummary}. Together with the published results
\cite{Reference:pdg2001,Reference:selexd0lt,Reference:selexdslt} 
I have included the 
preliminary results shown
in the figure in my new world averages given in
Table \ref{tb_worldaverages}. For the $D_s^+$ FOCUS preliminary
result \cite{Reference:focusdslt}\ which does not yet include 
a systematic uncertainty I have taken
the total uncertainty to be $\sqrt{2}$ times the statistical error.
The fixed-target and $e^+e^-$ averages are also separately shown and
agree fairly well suggesting no additional unaccounted-for systematics.
The $e^+e^-$ averages are currently dominated by the preliminary BELLE
measurements.

\begin{table}
\begin{tabular}{lc|lc}
\hline
  \tablehead{1}{l}{b}{Quantity} 
  & \tablehead{1}{c}{b}{World Average}
  & \tablehead{1}{l}{b}{Quantity} 
  & \tablehead{1}{c}{b}{World Average} \\ \hline
$\tau(D^+)$ (fs)              & $1044.1\pm 9.1$ & 
$\tau(\Lambda_c^+)$ (fs)      &  $200.2\pm 3.2$ \\
$\tau(D^0)$ (fs)              &  $413.3\pm 1.7$ & 
$\tau(\Xi_c^+)$ (fs)          &  $433\pm 19$ \\
$\tau(D_s^+)$ (fs)            &  $492.0\pm 5.3$ & 
$\tau(\Xi_c^0)$ (fs)          &  $106.0^{+9}_{-8}$ \\
\hline
\end{tabular}
\caption{World averaged lifetimes including also preliminary results.}
\label{tb_worldaverages}
\end{table}

The ratio $\tau(D_s^+)/\tau(D^0)$ continues to be of interest in
determining the importance of the suppression of W-exhange and W-annihilation
contributions in $D$-meson decays. There are now three direct measurements
of this ratio giving an average of $1.171\pm 0.018$ which agrees well with
just taking the ratio of the two world average lifetimes:
$1.190\pm 0.014$. The ratio is much larger than $1.07$ which is the maximum
expected size for no W-exchange/W-annihilation contributions
\cite{Reference:bigiuraltsev94}. An accurately measured value for this
ratio can be used to determine phenomenlogically the relative size of
W-exchange/W-annihilation contributions \cite{Reference:cheung99}.

\begin{figure}
  \includegraphics[height=1.5in,width=.33\textwidth]{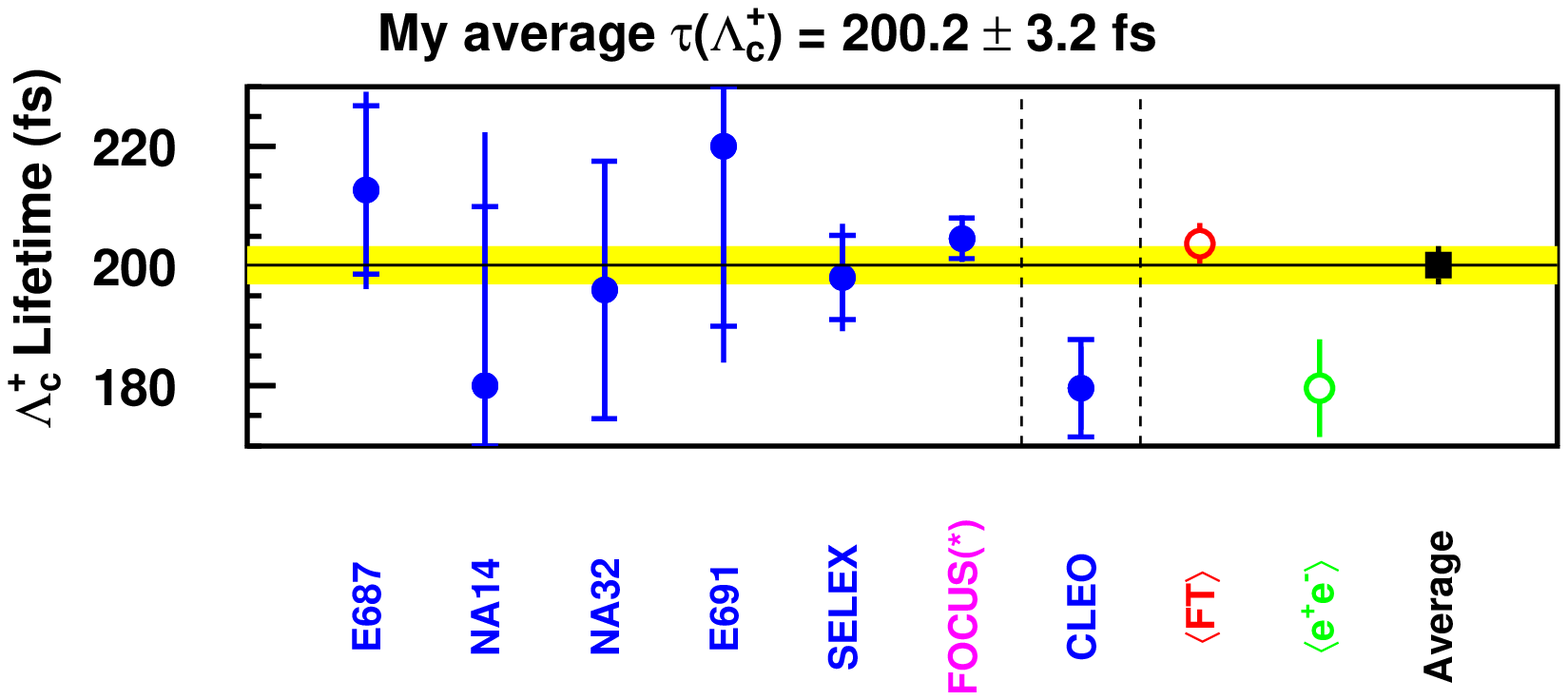}
  \includegraphics[height=1.5in,width=.33\textwidth]{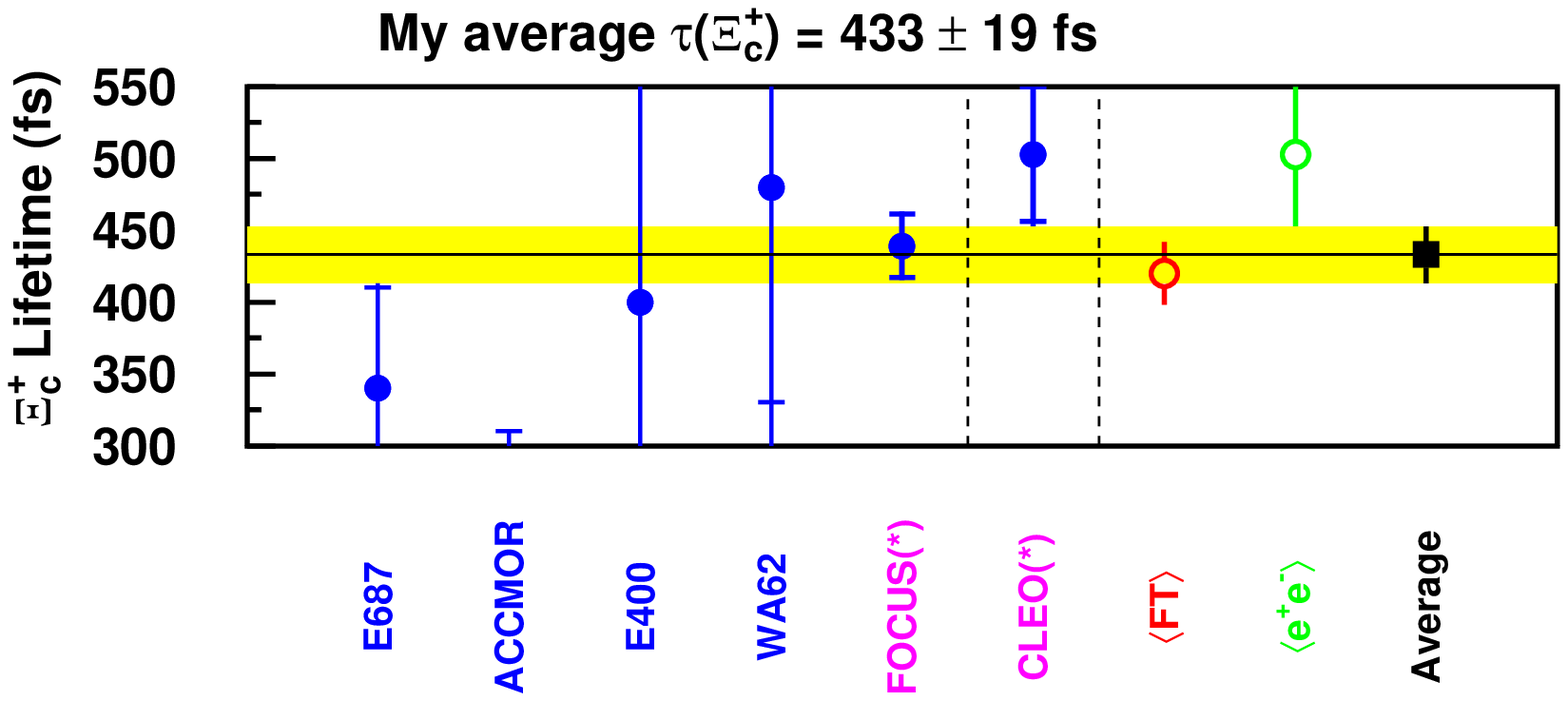}
  \includegraphics[height=1.5in,width=.33\textwidth]{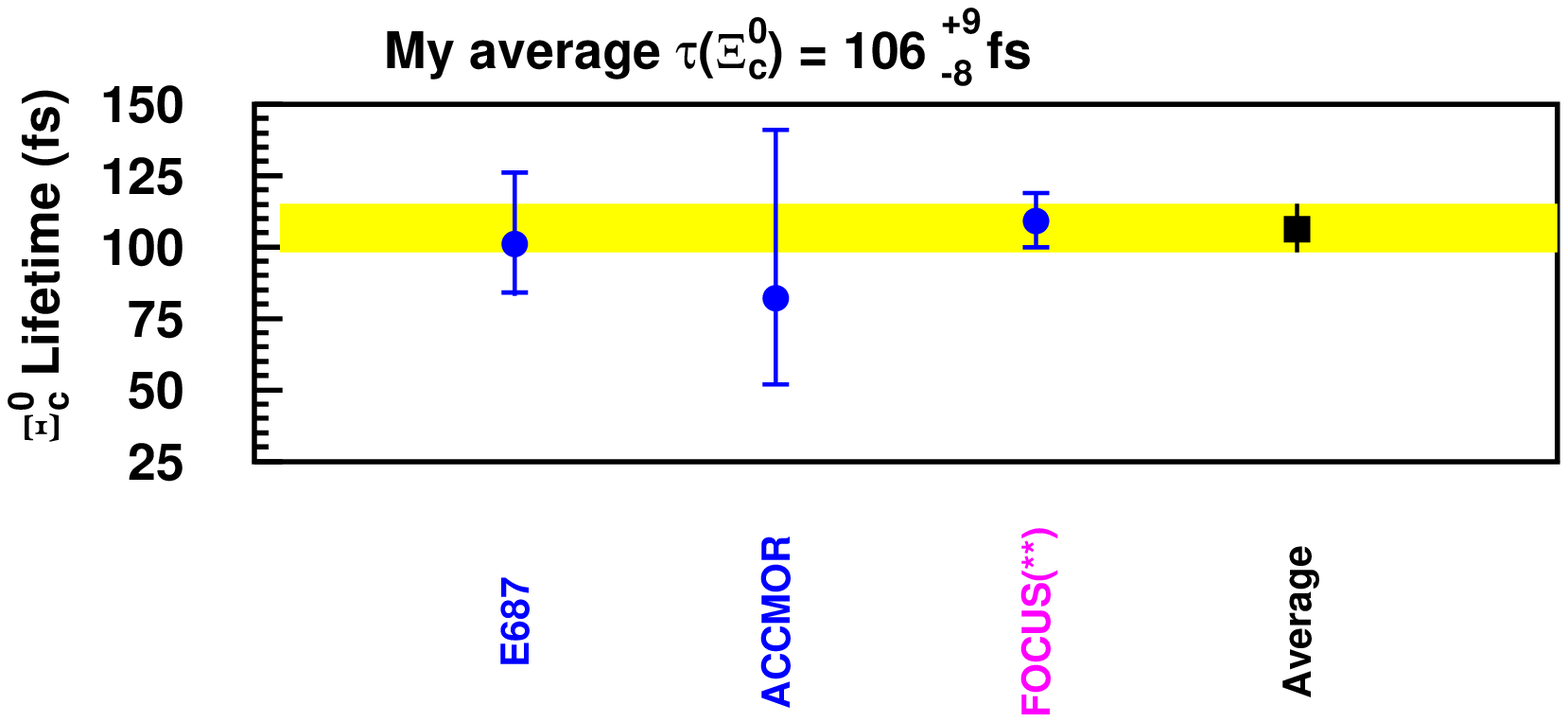}
  \caption{Summaries of recent charm baryon lifetime measurements
and their averages.}
\label{fg_barltsummary}
\end{figure}

There are new preliminary FOCUS lifetime results for the
$\Lambda_c^+$, $\Xi_c^+$ and 
$\Xi_c^0$ baryons \cite{Reference:eric2001,Reference:eduardo2001}
and a CLEO preliminary
lifetime result for $\Xi_c^+$ \cite{Reference:cleoxicpplt}.
These are shown together with published results in
Fig.\ref{fg_barltsummary}. As previously, for the preliminary FOCUS
lifetime result for $\Xi_c^0$ which does not include a systematic
uncertainty I have taken the total uncertainty to be $\sqrt{2}$ times
the statistical error in determining the world average.

There are two items of note in the new results. The first is that the
CLEO published lifetime for $\Lambda_c^+$ appears to disagree with the
fixed-target average value when the new FOCUS preliminary number is
included. This could point to a systematic problem starting to appear
which might be related to the short-lived nature of the $\Lambda_c^+$
decay. An example of a possible effect is the mass-lifetime correlation
seen in CLEO which introduces very large MC correlations to the lifetime.
It is known that the size and type of the resonance sub-structure of a decay 
like $\Lambda_c^+\rightarrow pK^-\pi^+$ can be important to the
acceptance corrections, and may also be important in the mass-lifetime
correlations since the resonance sub-structure alters the angular 
distributions of the daughter tracks. Uncertainties related to the
resonance sub-structure can thus introduce additional systematic
studies. Any systematic problems between the fixed-target and
$e^+e^-$ results should be kept in mind as the BaBar 
and BELLE lifetime and mixing results become more precise.

The other interesting feature of the new results is that the ratio
$\tau(\Xi_c^+)/\tau(\Lambda_c^+)$ is now $2.16\pm 0.11$ compared to the
PDG2000 value of $1.60^{+0.31}_{-0.22}$. Many of the calculations favour a
value of $1.2$--$1.6$ though with large 
uncertainties \cite{Reference:ltcalculations}. 
It would be interesting
to see if one can feed back into the calculations this new information
and others like the $\tau(D_s^+)/\tau(D^0)$ ratio to get a better
understanding, and to see if it affects the prediction for other related
quantities like the ratio $\tau(\Lambda_b^0)/\tau(B^0)$. Both the
mentioned results suggest that the W-exchange contribution 
is much more important than have so far been assumed in the calculations.

\section{D-Mixing Review}

The parameters we use to describe D-mixing can best be defined by the
relevant equations relating the states with definite mass and lifetimes:
$\vert D_{H}(t)\rangle=e^{-im_{H}t}e^{-\Gamma_Ht/2}\vert D_{H}\rangle$, and
$\vert D_{L}(t)\rangle=e^{-im_{L}t}e^{-\Gamma_Lt/2}\vert D_{L}\rangle$ 
to the observed
$\vert D^0\rangle$ and $\vert \overline{D^0}\rangle$ states: 
$\vert D_{H}\rangle=p\vert D^0\rangle + q\vert \overline{D^0}\rangle$ and
$\vert D_{L}\rangle=p\vert D^0\rangle - q\vert \overline{D^0}\rangle$. 
So for example the
doubly-Cabibbo-suppressed (DCS) decay rate 
$\Gamma(D^0\rightarrow K^+\pi^-)=
\vert \langle K^+\pi^-\vert T\vert D^{0}(t)\rangle\vert^2$.
We will assume CP conservation in charm decays and use the following
approximations for charm: $\vert q/p\vert=1$ and 
$\vert x\vert, \vert y\vert, R_{DCS} << 1$. Where
$x={\Delta m/\Gamma}$, $\Delta m=m_H-m_L$,
$\Gamma=\left(\Gamma_H+\Gamma_L\right)/2$,
$y={\Delta\Gamma/2\Gamma}=
{(\Gamma_{CPeven}-\Gamma_{CPodd})/(\Gamma_{CPeven}+\Gamma_{CPodd})}
=y_{CP}$,
and $R_{DCS}=\vert{\langle K^+\pi^-\vert T\vert D^{0}\rangle/
\langle K^+\pi^-\vert T\vert \overline{D^{0}}\rangle}\vert^2$.
The ratio of the ``wrong-sign'' $D^0\rightarrow K^+\pi^-$ to
``right-sign'' $D^0\rightarrow K^-\pi^+$ decays is given by
$$R_{WS}(t)=\left[R_{DCS}+(y{\rm cos}\delta - x{\rm sin}\delta)t\sqrt{R_{DCS}}+
{(x^2+y^2)\over 4}t^2\right]e^{-t}$$
$$R_{WS}(t)=\left[R_{DCS}+y^{\prime}t\sqrt{R_{DCS}}+
{(x^{\prime 2}+y^{\prime 2})\over 4}t^2\right]e^{-t}$$
where $y^{\prime}\equiv y{\rm cos}\delta - x{\rm sin}\delta$ and
$x^{\prime}\equiv x{\rm cos}\delta + y{\rm sin}\delta$ and
$\delta$ is the strong phase difference between the Cabibbo-favoured and
the DCS decay.

Information on the charm mixing parameters can be obtained in several ways:
\begin{enumerate}
\item
Measure the lifetime difference between CP-even, CP-odd and flavour
specific states to give $y_{CP}$.
\item
Measure wrong-sign semileptonic decays which do not require good lifetime
resolution but only give information on $(x^2+y^2)$ and cannot separate
$x$ and $y$.
\item
Measure wrong-sign hadronic decays like $D^0\rightarrow K^+\pi^-$ which
require a lifetime study and high S/B and can give information on both
$x^2$ and $y$ separately. However there is an additional complication of an
unknown strong phase difference, e.g. $\delta_{K\pi}$ 
between $D^0\rightarrow K^+\pi^-$
and $D^0\rightarrow K^-\pi^+$ contributions.
\end{enumerate}

\begin{figure}
  \includegraphics[height=1.5in]{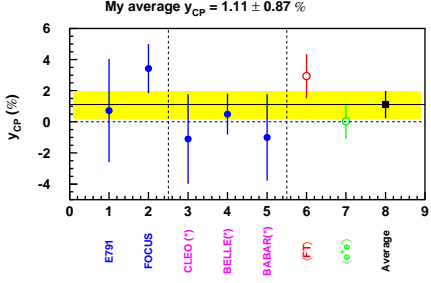}
  \caption{Summary of recent $y_{CP}$ measurements.}
\label{fg_ycpsummary}
\end{figure}

\begin{figure}
  \includegraphics[height=2.5in]{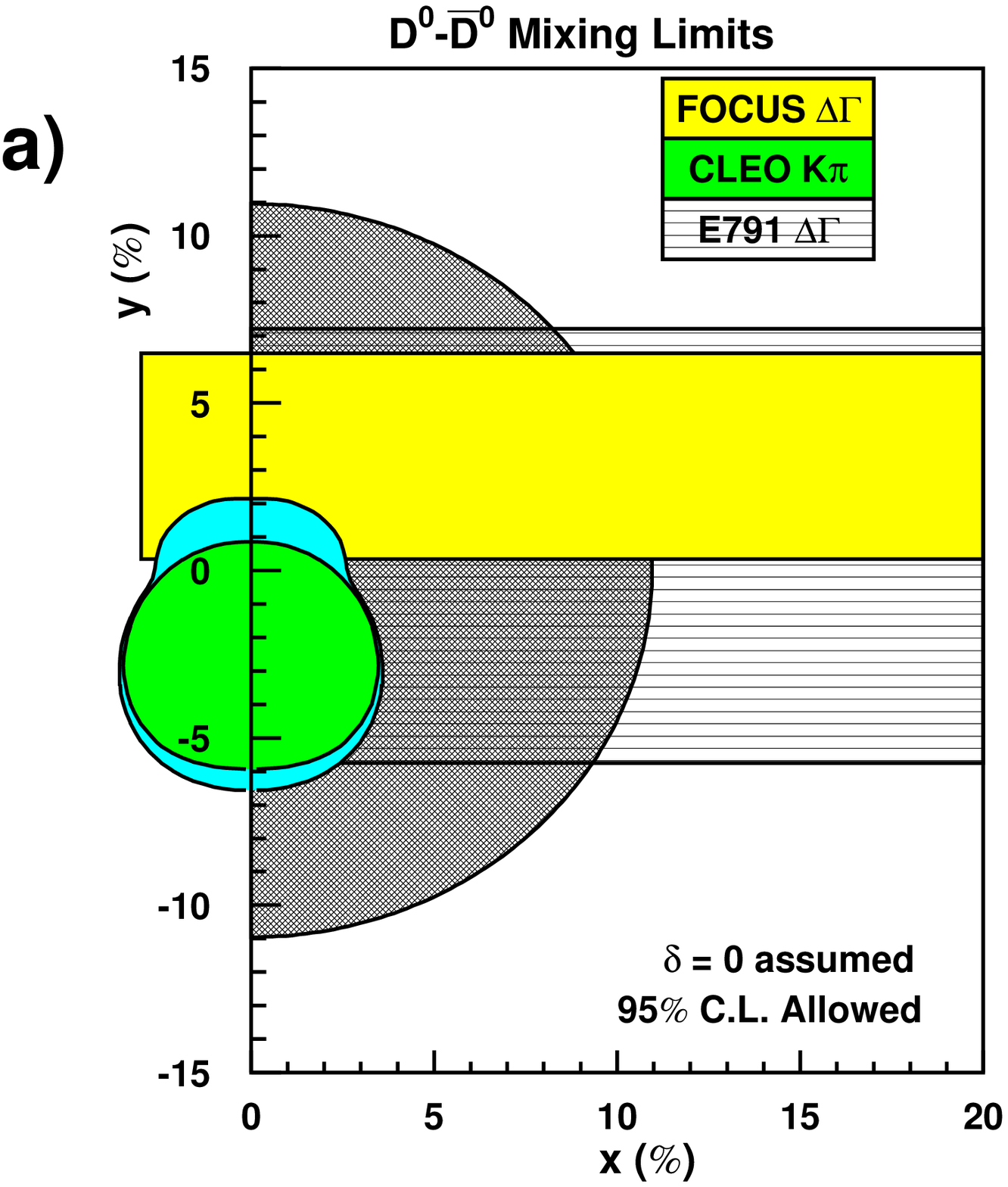}
  \includegraphics[height=2.5in]{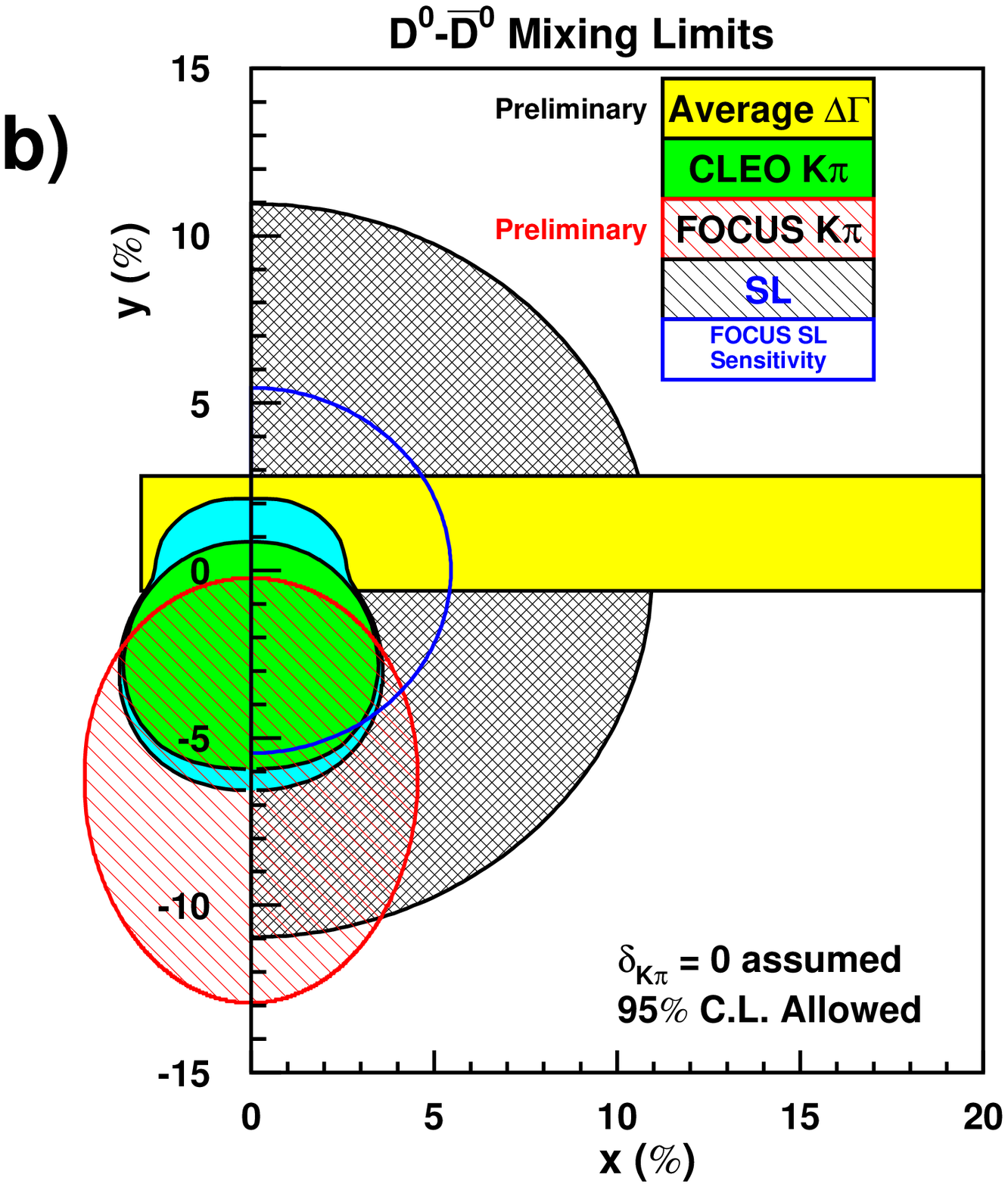}
  \includegraphics[height=2.5in]{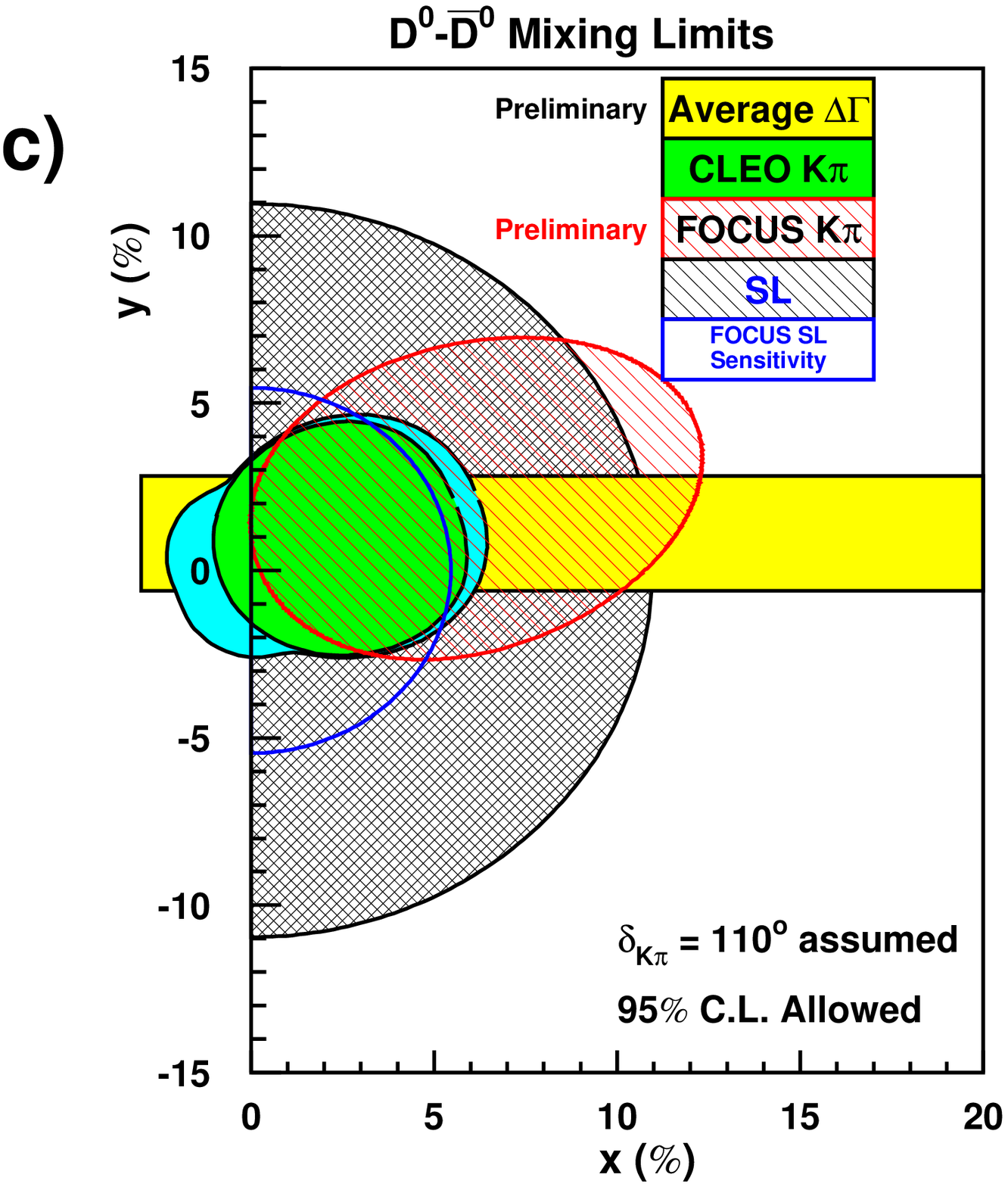}
  \caption{Summary of D-mixing results showing 95\%\ confidence level
allowed regions in ($x$,$y$) (a) for published results assuming 
$\delta_{K\pi}=0$; and including preliminary results assuming
(b) $\delta_{K\pi}=0$ and (c) $\delta_{K\pi}=110^{\circ}$.}
\label{fg_mixsummary}
\end{figure}

The published results are summarized in Fig.\ref{fg_mixsummary}(a)
which include the $y_{CP}$ measurement from FOCUS \cite{Reference:focusycp},
the limits on $(x^{\prime}$,$y^{\prime})$ from CLEO allowing and not
allowing for CP violation \cite{Reference:cleodmix}, and the
E791 limits from semileptonic decays \cite{Reference:e791sl}.

The preliminary measurements from CLEO \cite{Reference:cleoycp}, 
BELLE  \cite{Reference:bellelt}\ and 
BaBar  \cite{Reference:babarycp}
for $y_{CP}$
are given in Fig.\ref{fg_ycpsummary}. Including these produces a
world average of $1.11\pm 0.87$\%\ which is quite consistent with 
zero\footnote{At ``press time'' the $y_{CP}$ value of
$+0.5\pm1.0^{+0.8}_{-0.9}$ shown by BELLE at the conference was
superceeded by a new value of $-0.5\pm1.0^{+0.7}_{-0.8}$ contained
in their new preprint \cite{Reference:belleycp}. The same data
sample was used but the analysis contained updated MC corrections. 
This moves the world average
value down to $0.63\pm 0.85$.}.
However the situation is still interesting due to in part to
a preliminary result
from FOCUS for the allowed region in $x^{\prime}$ and $y^{\prime}$
\cite{Reference:link2001}. The 95\%\ confidence level allowed regions
for the FOCUS and CLEO measurements are given in 
Fig.\ref{fg_mixsummary}(b) in the ($x$,$y$) space assuming the strong phase
difference $\delta_{K\pi}=0$. The slightly larger (lighter) region for CLEO
is when CP-conservation is not assumed.
Also shown are the allowed region in $y$ from the combined lifetime difference
measurements assuming no CP-violation, and the circular allowed region from
wrong-sign semileptonic decay limits from E791 \cite{Reference:e791sl}. 
The smaller circular line gives the
expected size of the allowed region from FOCUS wrong-sign semileptonic 
decays \cite{Reference:focussl}.

Even though I do not have the likelihood contour of the CLEO allowed region
in order to combine the FOCUS and CLEO results,
it can be seen that the combined allowed region is beginning to 
exclude zero, especially if no CP-violation is assumed. The other point is
that the combined FOCUS and CLEO $K\pi$ result for the allowed region does not
agree well with the world average allowed $y_{CP}$ range when one assumes
$\delta_{K\pi}=0$. This could indicate one of three things: it is just a
statistical fluctuation; the systematic uncertainties are underestimated;
or the most interesting is that $\delta_{K\pi}$ is large and non-zero.
Knowing the value of $\delta_{K\pi}$ is crucially important. For example
in the absence of a theoretically favoured value,
the experimentally preferred value is about
$110^{\circ}$ as shown in Fig.\ref{fg_mixsummary}(c). This would be a
really interesting situation since the favoured scenario is with $y$
near zero and a large value of $x$ of $\approx 3$\%\footnote{Note that the
new lower world average value of $y_{CP}$ does not significantly change these
favoured values.}. This is the most likely expected
signature of new physics beyond the Standard Model since these
can produce
sizable non-zero values in $\Delta m$ but not usually in $\Delta\Gamma$.

\section{Conclusions}

Charm lifetime measurements continue to be an interesting way to study
non-perturbative strong interaction physics and evaluate possible systematic
problems for measurements that require good lifetime resolutions. The
larger than expected values of $\tau(D_s^+)/\tau(D^0)$ and
$\tau(\Xi_c^+)/\tau(\Lambda_c^+)$ indicates that W-exchange is much more
important that normally thought which could have implications on other
theoretical predictions, like $\tau(\Lambda_b^+)/\tau(B^0)$.

The D-mixing situation is still very interesting as the data now favour
a $y$ value near zero and a large value of $x$ of $\approx 3$\%. This could
be a signature of new physics beyond the Standard Model. However
its requires the strong phase difference $\delta_{K\pi}\approx 110^{\circ}$
which is unexpected theoretically. The possibility that this is a
statistical fluctuation or a systematic underestimation can be greatly
clarified by new precise D-mixing results from BaBar and BELLE and we
look forward eagerly for these results in the future.

\begin{theacknowledgments}
I wish to thank Monika Grothe and Jun'ichi Tanaka for providing the
BaBar and BELLE results respectively, and for providing prompt replies
to my questions.
\end{theacknowledgments}




\end{document}